# Metamorphism: A Mathematical Challenge for Antivirus Technology*

Luis M. Augusto

October 21, 2026

Independent Researcher

https://orcid.org/0000-0001-9097-5722

luis.ml.augusto@protonmail.com

**Abstract**

*Metamorphic viruses, currently the most advanced computer viruses in the wild, have the unique ability of mutating their own code virtually into infinitely many highly dissimilar copies of themselves that nevertheless have the same functionality. This ability—metamorphism—together with other advanced obfuscation techniques makes these computer viruses virtually undetectable by the antivirus technology available in the market today. In this paper, I show that viral metamorphism is fully attainable by the employment of van Wijngaarden grammars. The challenge then for an antivirus software is to embed the Turing machine that decides the language generated by the grammar.*

## 1 Introduction

Computer viruses, which have been around since the early 1970s, are essentially computer programs that can replicate their own code in their infected targets. Their growth in number and type has been exponential from early on: Tellingly, in less than five years M. A. Ludwig had to revise the title of his book first published in 1991 by changing it from *The **Little** Black Book of Computer Viruses* [1] to *The **Giant** Black Book of Computer Viruses* [2] in 1995—my bold-font emphasis. But then things got much, much worse when antivirus software companies had to worry not only about the number of viruses in the wild but also with the mathematical properties of a few of them. In effect, in the early 2000s some clever computer virus designers realized the malicious potential hidden in mathematics. This general approach to the design of malware, now known as *malicious mathematics* and generally explored in a few monographs (e.g., [3,4]) and papers (e.g., [5,6]), forced antivirus software companies to upgrade their staff and technology to the highest level in computing.[1]

*Published as a working paper in Ω ::=*Journal of Formal Languages* under the title "Metamorphism: A mathematical challenge for antivirus software"; submitted for standard publication until acceptance.

[1]Malicious mathematics finds applications in other fields where malicious intent poses a constant threat, such as in cryptography; see, e.g., [7].

As a matter of fact, not all software companies bothered to do so, as only—possibly negligibly—few virus designers can cope with the hard requirements of mathematics, malicious or otherwise. This is a particularly Herculean task in the case of the metamorphic viruses, i.e. viruses that generate highly dissimilar replicas of themselves to escape detection and analysis by antivirus technology (AVT); see, e.g., [8-12]. Every computer virus is a program written in some computer language such as an assembly language or a high-level programming language; as such, writing a "typical" virus requires from a virus designer no more maths than writing a typical program does from a programmer. In fact, your usual "lamers" or "script kiddies" (i.e. unskilled malware users) can easily, albeit illegally, download kits available on the Internet that basically write the viruses for them. In contrast, metamorphic viruses require, besides the usual computing expertise, strong mathematical skills both in formal languages and automata theory and in computability theory, two subjects that are typically conjoined in a single course in a computer science curriculum (e.g., [13]).[2]

Thus, writing a successful metamorphic virus is a computational feat, so much so that most antivirus companies regard the threat of one such virus as abstract as the mathematics it is created on. (An honorable exception is Symantec, which as early as 2003 already included these viruses in its white paper; cf. [15].) But the payoff of going to the great lengths of writing a metamorphic virus is their undetectability, with polymorphic techniques, less advanced than metamorphism, already placing a virus in the **NP**-complete class [16]. In other words, the payload for these virtually undetectable viruses may belie the old motto that "crime does not pay." These viruses are evolving slowly, but their hybridization with advanced obfuscation techniques such as $k$-ary coding simply cannot be ignored; see, e.g., [17] and [18]. Additionally, their predictably core role in future warfare makes their comprehensive analysis not only necessary, but also urgent.

Here, drawing on F. Cohen's original formalization of computer virology based on the Turing machine I associate metamorphic viruses with a special kind of formal grammars, to wit, the van Wijngaarden grammars. I claim that a *pure metamorphic virus* can be designed entirely by means of a van Wijngaarden grammar; the challenge then for an AVT is to embed the Turing machine that decides the language generated by the given grammar. The present discussion continues previous work in [19-21] on the syntax and semantics of these grammars and sets a formal research program in computer virology.

# 2 Preliminary Definitions and Results

## 2.1 The Turing Machine and the Formal Definition of a Computer Virus

### 2.1.1 Computable and partial computable functions

A computer virus (henceforth often just "virus") is essentially a computer program capable of replicating its own code in a host program or file and possibly evolving different copies of itself in this act. This process, in which host programs and files are modified, can be compared to a viral infection in the biological realm; hence the coinage of these specific programs. The payload of a virus is typically—albeit not necessarily—malicious: Causing denial of service or even system collapse, compromising data confidentiality or integrity, and extortion and blackmail are among the most frequent payloads as mentioned in, for example, [22]. These pernicious results typically take dramatic proportions when the targets are so-called critical systems, not only causing substantial financial damage but also endangering human lives. For this reason, virus designers take great care that their creations are not easily detected by the many antivirus software available in the market and to achieve this goal they are willing to go to great lengths, including the employment of advanced mathematical skills. In effect, all these aspects, to wit, *self-replication*, *evolution*, and *obfuscation* are mathematically well-founded. The most successful virus is, however, the one that simply cannot be identified as viral code, a result that only *undecidability* can provide. All these aspects emphasized in italics find their mathematical support in *computability theory*.[3]

[2]The first published paper in which this employment of maths in writing metamorphic viruses was elaborated on was [14].

In this Section, I elaborate on these mathematical results. The core result is the canonical definition of computable function:[4]

**Definition 1.** A function is *computable* if and only if (abbr.: iff) it is *effectively calculable*, i.e. iff there is an algorithm that calculates it.

A special group of computable functions are the primitive computable functions:

**Definition 2.** A function $f$ over the natural numbers is said to be *primitive computable* if it can be generated from the functions $z(x) = 0$, $succ(x) = x + 1$, and $U_i^n(\overline{x}) = x_i$—known as zero function, successor function, and projection function, respectively—by zero or more applications of the following schemata known as the *composition* or *substitution schema* (1) and the *primitive recursive schema* (2):[5]

$$(1) \qquad f(\overline{x}) = g(h_1(\overline{x}), ..., h_m(\overline{x}))$$

$$(2) \qquad \begin{cases} f(0, \overline{x}) = f(\overline{x}) \\ f(y + 1, \overline{x}) = f(y, f(y, \overline{x}), \overline{x}) \end{cases}$$

If

$$(3) \qquad f(\overline{x}) = \mu y R(\overline{x}, y)$$

known as *minimalization schema* and where $R$ is a relation symbol, is also applied then the function obtained is *partial computable*.[6]

I can now refine Definition 1 above as follows:

**Definition 3.** A function $f$ is *computable* iff it is total and *effectively calculable*; it is *partial computable* iff it is effectively calculable.

The primitive computable functions allow for an enumeration of the partial computable functions:

**Theorem 4.** *There are exactly countably infinitely many partial computable functions and there are exactly countably infinitely many computable functions.*

Recall here that "countably infinitely many" is denoted by $\aleph_0 = \omega = |\mathbb{N}|$, where $\mathbb{N} = \{0, 1, 2, ...\}$ and for a set $S$ the notation $|S|$ denotes the cardinality of $S$.

[3] Also often, but less and less frequently, called *recursion theory*. For this synonymy-loss phenomenon, see [23,24].

[4] I assume the reader has basic knowledge of mathematical objects such as functions and sets; the reader is referred to [13] for these foundational objects and for the essential aspects of computability theory, which include a formal discussion on the Turing machine.

[5] Recall that we write "$f^n(\overline{x})$" as an abbreviation for "$f(x_1, ..., x_n)$". In practice, we often write simply $f(\overline{x})$ if the arity of $f$ is obvious or of no import. It is right to speak of *recursiveness* w.r.t. (2), but we can also say that (2) is the primitive *computable* schema.

[6] Intuitively, the $\mu$-operator is a search operation in the sense that $f(\overline{x})$ = the least $y$ s.t. $g(\overline{x}, y) = 0$ for some $g$. If no such $y$ exists, then the search may go on for ever and $f(\overline{x})$ never gets defined. Hence the inclusion of partial computability.

#### 2.1.2 Turing machines with temporal functions and computer viruses

Recall that function $f$ is written $\varphi$ if it is a partial function, i.e. if it is undefined for some element(s) in its domain. As usually, my model of effective calculability or computability will be the Turing machine in the following sense known as *Church-Turing thesis* or *Turing's theorem* (cf. [25]):

**Theorem 5.** *A function that is effectively calculable is a function that can be computed by a Turing machine.*

**Definition 6.** I define a Turing machine to be the 6-tuple $\mathscr{M} = (Q, \Sigma, D, \varpi, \delta, \partial)$ where $Q = \{q_0, q_1, ..., q_k\}$ is the set of $k+1$ states with $q_0$ the initial state, $\Sigma = \{a_1, ..., a_n\}$ is the $n$-letter machine alphabet, $D = \{l, s, r\}$ is the set of the possible tape motions *left*, *stop*, or *right*, $\varpi : (Q \times \Sigma) \longrightarrow \Sigma$ is the output function, $\delta : (Q \times \Sigma) \longrightarrow Q$ is the state-transition function, and $\partial : (Q \times \Sigma) \longrightarrow D$ is the motion function.

For my purposes, with F. Cohen [26,27] I define a Turing machine performing temporal functions in the following way:

**Definition 7.** Let $\mathscr{M}$ be a Turing machine. Let the tape of $\mathscr{M}$ be the set $C_{\mathscr{M}} = \{c_i\}_{i=0}^{\omega}$ where $c_i \in C_{\mathscr{M}}$ denotes the $i$-th tape cell s.t. $c_0$ is the leftmost tape cell and $c_i^{\#}$ is the $i$-th empty cell. Further define the set of steps or moves of a Turing machine to be $S_{\mathscr{M}} = \{s_j\}_{j=0}^{\omega}$ for $s_j \in S_{\mathscr{M}}$ the $j$-th step taken by $\mathscr{M}$ and s.t. $s_1$ is the first step taken by $\mathscr{M}$. Let now the notion of time *instant*, denoted by $t$ and s.t. $t'$ is the successor instant of $t$ if $t' > t$, coincide with the notion of *step* or *move*. Define now the following *temporal functions* performed by $\mathscr{M}$:

1. $\sigma : S_{\mathscr{M}} \longrightarrow Q$
2. $\zeta : (S_{\mathscr{M}} \times C_{\mathscr{M}}) \longrightarrow \Sigma$
3. $\varrho : S_{\mathscr{M}} \longrightarrow C_{\mathscr{M}}$

With the function triple $(\sigma, \zeta, \varrho)$ Cohen defines the history of $\mathscr{M}$ at time $t$ as

$$H_{\mathscr{M}}(t) = \sigma(t), \zeta(t, c), \varrho(t)$$

and s.t.

$$\mathcal{H}_{\mathscr{M}} = \bigcup_{i=0}^{k} H_{\mathscr{M}}(t_i)$$

where $H_{\mathscr{M}}(t_0)$ is the initial state of $\mathscr{M}$, is the complete history—i.e. the *computation*—of $\mathscr{M}$.

Note in this definition that we consider a Turing machine computation to be complete iff it is performed in finite time in at least one step or move, i.e. $0 < i \leq k$ for some $k \in \mathbb{N}$.

**Definition 8.** A Turing machine $\mathscr{M}$ halts at instant $t$ iff

$$(\exists t)(\forall t' > t)[\sigma(t) = \sigma(t')]$$

$$(\exists t, t')(\forall i \in \mathbb{N})[\zeta(t, c_i) = \zeta(t', c_i)]$$

and

$$(\exists t, t')[\varrho(t) = \varrho(t')].$$

Note that Definition 8 employs the language of first-order predicate logic.[7] Let now $v \in \Sigma^i$, $i = 0, 1, 2, ...$, be a word of length $i$ over the alphabet $\Sigma$ s.t. $\Sigma^0 = \{\lambda\}$, $\Sigma^1 = (v_1 = a_j)$, $\Sigma^2 = (v_1 v_2 = a_{j,1} a_{j,2})$, etc., where $a_j$ is the $j$-th letter in $\Sigma$. We define $\Sigma^* = \bigcup_{i=0}^{k} \Sigma^i$ and $\Sigma^+ = \bigcup_{i=1}^{k} \Sigma^i$, where for practical ends we limit $i$ to some $k \in \mathbb{N}$.

**Definition 9.** A Turing machine program is a finite sequence of letters from $\Sigma$, called a *string*, constructed as

$$(\psi) \qquad \left(\zeta\left(t, c_j\right), ..., \zeta\left(t, c_{j+|v|-1}\right)\right) = v$$

if

$$(\forall \mathscr{M} \in \mathcal{M}) (\forall v) (\forall i \in \mathbb{N}) \left[v \in \widehat{\mathscr{M}} \Leftrightarrow v \in \Sigma^*\right]$$

where $\mathcal{M}$ is the countable set of all the Turing machines and $\widehat{\mathscr{M}}$ denotes the program of Turing machine $\mathscr{M}$.

*Remark* 10. More strictly defined, we have $\mathcal{M} = \{\mathscr{M}_e\}_{e=0}^{\omega}$, where $e \in \mathbb{N}$ is an *index*, i.e. every Turing machine has at least a *code number* that unequivocally describes it. This code number is typically the *Gödel number* of a given Turing machine.[8]

We now prevent the existence of an empty set of Turing machine programs in the following way:

**Definition 11.** The set $\widetilde{\mathcal{M}}$ s.t. $\left|\widetilde{\mathcal{M}}\right| \neq \emptyset$ is formally defined as:

$$(\forall \mathscr{M} \in \mathcal{M}) (\forall V) \left[V \in \widetilde{\mathcal{M}} \Leftrightarrow \left(\exists v \in V \wedge \forall v \in V \left(v \in \widehat{\mathscr{M}}\right)\right)\right]$$

Employing the language of second-order logic, we define mathematically a computer virus in the following way:[9]

**Definition 12.** Let $\mathscr{M}$ denote a Turing machine as defined above and let us identify a Turing machine $\mathscr{M}$ with its program s.t. $\widehat{\mathscr{M}} = \mathscr{M}$. A computer virus, denoted by $v \in V$, is defined as:

$$(\forall \mathscr{M}) (\forall V) ((\mathscr{M}, V) \in \mathcal{V} \Leftrightarrow (\left(\left(V \in \widetilde{\mathcal{M}}\right) \wedge (\mathscr{M} \in \mathcal{M}) \wedge\right.$$

$$(\forall v \in V) (\forall H_{\mathscr{M}}) \left((\forall t) (\forall j) \left[\underbrace{(\varrho(t) = c_j)}_{\varphi_1} \wedge \underbrace{(\sigma(t) = \sigma(q_0))}_{\varphi_2}\right]\right) \wedge \psi) \Rightarrow$$

$$((\exists v' \in V) (\exists t' > t) (\exists j') \underbrace{([(j' + |v'|) \leq j] \vee [(j + |v|) \leq j'])}_{\varphi_3} \wedge \psi' \wedge$$

$$\left(\underbrace{(\exists t'') (t < t'' < t') [\varrho(t'') \in j', ..., j' + |v'| - 1]}_{\varphi_4}\right))))$$

where $\psi$ is as in Definition 9 above and $\psi'$ is the string $\zeta(t', c_{j'}), ..., \zeta\left(t', c_{j'+|v'|-1}\right) = v'$.

[7] See, e.g., [28] for this language.

[8] We set it so that $\mathscr{M}_0$ is the empty Turing machine. See any of the cited texts on computability for the calculation of Gödel numbers for Turing machines.

[9] See [28] for the basic aspects of this language.

This definition, where $\mathcal{V}$ denotes the set of all viral sets, states that the pair $(\mathscr{M}, V)$ is a viral set iff for each virus $v \in V$ and for all the computations of Turing machine $\mathscr{M}$ we have, for all instants $t$ and cells $c_j \in C_{\mathscr{M}}$, if $(\varphi_1)$ the tape head is in front of cell $c_j$ at $t$, and $(\varphi_2)$ $\mathscr{M}$ is in its initial state at $t$, and $(\psi)$ the tape cells starting at cell $c_j \in C_{\mathscr{M}}$ and ending at cell $c_{j+|v|-1} \in C_{\mathscr{M}}$ hold virus $v$, then there exists a virus $v'$ at time instant $t' > t$ and at cell $c_{j'}$ s.t. $(\varphi_3)$ cell $c_{j'}$ is far enough from position $v$, $(\psi')$ the tape cells starting at $c_{j'}$ hold virus $v'$, and $(\varphi_4)$ at some time instant $t''$ s.t. $t < t'' < t'$ virus $v'$ is written by $\mathscr{M}$.

## 2.2 Self-Replication and Kleene's Fixpoint Theorem

The formalization in Definition 12 can be abbreviated as

$$(\forall \mathscr{M})(\forall V)\left((\mathscr{M}, V) \in \mathcal{V} \Leftrightarrow \left(\left(V \in \widetilde{\mathcal{M}}\right) \wedge (\mathscr{M} \in \mathcal{M}) \wedge \underbrace{\left((\forall v \in V)\left[v \underset{\mathscr{M}}{\Longrightarrow} V\right]\right)}_{\chi}\right)\right)$$

where $\chi$ denotes that Turing machine $\mathscr{M}$ *copies* any virus $v$ into set $V$.

The self-replicating ability of a virus is formally secured by *Kleene's recursion theorem*, also known as *Kleene's fixpoint theorem*:[10]

**Theorem 13.** *For any computable function $f(x)$, there is an $e \in \mathbb{N}$ such that $\varphi_e = \varphi_{f(e)}$.*

The proof of this theorem, whose meaning is that the partial computable functions are closed under fixpoint definitions, can be found in any standard work in computability theory (cf. cited literature). A good reference to understand this theorem in the framework of self-replicating machines—and hence of computer viruses—is [29], where it is shown that for any Turing machine $\mathscr{M} \in \mathcal{M}$ there is a partial function $\mathscr{M} : \mathbb{N} \longrightarrow \mathcal{R}$, for $\mathcal{R} = \{R_1, R_2, ...\}$ a countable class of objects called *representations*, s.t. $\mathscr{M}(x) = R_x$ where $R_x$ is thus the representation of $x$ by $\mathscr{M}$ (e.g., a string of 1s and 0s). Let now $\mathscr{D} \in \mathcal{M}$; the author shows that there is an integer $m$ s.t. for all inputs $\mathscr{M}_m$, where $m \in \mathbb{N}$ is the index or code number (the Gödel number) of $\mathscr{M}$, i.e. $\mathscr{M}$'s formal self-description in the guise of a fixed natural number, yields $\mathscr{D}(m)$ as output. Intuitively, $\mathscr{D}$ must be understood as a special Turing machine that, given any $x$, outputs a replica of the Turing machine with code number $x$. Then, the fixpoint theorem shows that there must be a Turing machine that on any input whatsoever outputs its own replica or *self-representation* by $\mathscr{M}_m(x) = \mathscr{D}(m)$.[11]

[10] Recall that if $f$ is a function $f : S \longrightarrow S$ with $S \subseteq \mathbb{N}$, then a fixpoint of $f$ is any point $x \in S$ such that $f(x) = x$.

[11] The proof runs as follows: We let $d$ be a code number for Turing machine $\mathscr{D}$. For any $x$, we let $\psi$ be the constant function whose output on any input is $h(d, x)$. As the instructions for $\psi$ will depend on $x$ we can find a computable function $f$ s.t.

$$(\forall y)\left[\varphi_{f(x)}(y) = h(d, x) = \psi_x(y)\right].$$

By applying the fixpoint theorem we get an integer $n$ s.t.

$$(\forall y)\left[\varphi_n(y) = \varphi_{f(n)}(y) = h(d, x)\right]$$

and hence

$$(\forall y)\left[\mathscr{D}_{\varphi_n}(y) = \mathscr{D}_{h(d,n)}\right].$$

But given that by the definition of $h$ we have $\mathscr{D}_{\varphi_n} = \mathscr{M}_{h(d,n)}$, by setting $m = h(d, n)$ we get

$$(\forall y)\left[\mathscr{M}_m(y) = \mathscr{D}(m)\right].$$

We can improve on the above results by modifying Kleene's fixpoint theorem in the following way:

**Theorem 14.** *For any computable function $f(x)$, there is an $e \in \mathbb{N}$ such that $\varphi_e \underset{c}{=} \varphi_{f(e)}$.*

In this reformulation, I add the detail—that amounts to a giant leap—that the partial computable functions $\varphi_e$ and $\varphi_{f(e)}$ are computably equivalent, a property that I denote by the symbol "$\underset{c}{=}$" and call *computable equivalence*. By writing "$\varphi_e \underset{c}{=} \varphi_{f(e)}$", I emphasize that $\varphi_e$ and $\varphi_{f(e)}$ yield the same output because the Turing machines with Gödel numbers $e$ and $f(e)$ compute the same function, whereas the notation "$\varphi_e = \varphi_{f(e)}$" merely denotes that $\varphi_e$ and $\varphi_{f(e)}$ yield the same value.

## 2.3 Viral Evolution: Cohen's Largest Viral Set

Any virus is by definition capable of self-replication:

**Proposition 15.** *For all $\mathscr{M} \in \mathcal{M}$ and $V = \{v\}$, we have:*

$$((\mathscr{M}, \{v\}) \in \mathcal{V}) \qquad \Rightarrow \qquad \left(v \underset{\mathscr{M}}{\Longrightarrow} v\right)$$

The next best thing is *evolution*, i.e. the ability of a virus to modify its own code when replicating itself, so that in each replication $v'$ of $v$ we have it that $v'$ is a *mutation* of $v$. Cohen [26,27] defined this property in the following way:

**Definition 16.** A virus $v \in V$ is said to *evolve* into a virus $v' \in \Sigma^*$ if there is a Turing machine $\mathscr{M}$ s.t. the computation $v \underset{\mathscr{M}}{\Longrightarrow} \{v'\}$ is performed. More specifically, $v'$ is an evolution of $v$ iff, for the pair $(\mathscr{M}, V) \in \mathcal{V}$, there are $i \in \mathbb{N}$ and $V' \subseteq V^i$ s.t. $v, v' \in V$ and for all $v_k \in V'$ it is the case that $v_k \underset{\mathscr{M}}{\Longrightarrow} v_{k+1}$ and there are $l, m \in \mathbb{N}$ s.t. $l < m$ and $v_l = v$ and $v_m = v'$.

**Theorem 17.** *For all $\mathscr{M} \in \mathcal{M}$ and all $\Gamma^* \subset 2^{\Sigma^*}$ we have for all $V \subseteq \Gamma^*$:*

$$(\mathscr{M}, V) \in \mathcal{V} \qquad \Rightarrow \qquad \left(\mathscr{M}, \bigcup \Gamma^*\right) \in \mathcal{V}$$

Intuitively, this theorem states that any union of (a finite number of viral sets) is also a viral set. If we next consider that there is a set $U \subset \Sigma^*$ s.t. $(\mathscr{M}, U) \in \mathcal{V}$ and for all $v \in V$ it is the case that $v \in U$, we have the following definition:

**Definition 18.** $U$ is called the *largest viral set w.r.t. $\mathscr{M}$*, denoted by $\mathcal{L}(\mathscr{M})$.

Then, the following result follows naturally from Theorem 17 and from the consideration that if $v \underset{\mathscr{M}}{\Longrightarrow} v'$, then $v' \in \mathcal{L}(\mathscr{M})$:

**Corollary 19.** *Let the pair $(\mathscr{M}, V) \in \mathcal{V}$ denote that $V$ is a viral set w.r.t. $\mathscr{M}$, a denotation that is abbreviated as $V(\mathscr{M})$. Then:*

$$\mathcal{L}(\mathscr{M}) = \bigcup_{i=1} V_i(\mathscr{M})$$

We can set the $V_i(\mathscr{M})$ s.t. $|V_i| = i$ where $i \in \mathbb{N}$ denotes the fixed number of evolved forms of a virus $v \in V_i$. Clearly, the definition of $\mathcal{L}(\mathscr{M})$ as the union of all the $V_i$ entails that at least one of the $V_i$, namely $V_1$ s.t. $|V_1| = 1$, is a *smallest viral set w.r.t.* $\mathscr{M}$, denoted by $\mathcal{S}(\mathscr{M})$. A virus $v \in V_1$ is a virus that does not evolve; formally, we have:

**Proposition 20.** *For all $\mathscr{M} \in \mathcal{M}$ and all $v \in \Sigma^*$ s.t. $(\mathscr{M}, V) \in \mathcal{V}$, we have:*

$$\left(v \underset{\mathscr{M}}{\Longrightarrow} \{v\}\right) \qquad \Rightarrow \qquad ((\mathscr{M}, \{v\}) \in \mathcal{V})$$

Compare this with Proposition 15 above. In the next Section, I explain the practical meaning of this proposition.

# 3 Evolving Viruses and Further Computability Theory

## 3.1 Back to the Turing Machine

With the formal definition of a virus at hand, from the viewpoint of computability theory three questions are now posed. In order both to pose them and to reply to them we simplify the halting function defined for a Turing machine $\mathscr{M}$ (cf. Def. 8):

**Definition 21.** Given a Turing machine $\mathscr{M}$, the function $halt_{\mathscr{M}} : T \longrightarrow Q_{\mathscr{M}}$, where $T = \{t_0, t_1, ...\}$ is a set of time instants and $Q_{\mathscr{M}}$ denotes the state set of Turing machine $\mathscr{M}$, is called the *halting function* for $\mathscr{M}$.

Clearly, the halting function for a Turing machine $\mathscr{M}$ has a value iff at a given time $t \in T$ machine $\mathscr{M}$ halts at state $q_i \in Q_{\mathscr{M}}$, i.e. iff we have $halt_{\mathscr{M}}(t) = q_i$. Otherwise, $\mathscr{M}$ does not halt; for instance, it loops forever. Recall from Definition 8 that in this case it is not possible to describe the complete computation performed by $\mathscr{M}$.

**Problem 22.** The three questions are as follows:

1. Given some code $v \in V \subseteq \Sigma^*$, is there a Turing machine $\mathscr{D} \in \mathcal{M}$ that can decide in finite time whether $v$ is a virus? Formally:

$$(\exists \mathscr{D})(\exists t)(\exists q_i)(\exists V)(\forall \mathscr{M})\left[(halt_{\mathscr{D}}(t) = q_i) \Leftrightarrow ((\mathscr{M}, V) \in \mathcal{V})\right]?$$

2. Given a pair $(v, v') \in V$, is there a Turing machine $\mathscr{D} \in \mathcal{M}$ that can decide in finite time whether $v'$ is an evolution of $v$? Formally:

$$(\exists \mathscr{D})(\exists t)(\exists q_i)(\exists v, v')(\forall \mathscr{M})\left[(halt_{\mathscr{D}}(t) = q_i) \Leftrightarrow \left(v \underset{\mathscr{M}}{\Longrightarrow} \{v'\}\right)\right]?$$

3. Given any string $x \in \{0,1\}^i$ for $i \in \mathbb{N}$, is there a particular class of Turing machines $\mathscr{M}'$ that can decide whether $x$ is a viral evolution? Formally:

$$(\forall \mathscr{M}')(\exists (\mathscr{M}, V) \in \mathcal{V})(\forall x)\left[(x \in \mathcal{H}_{\mathscr{M}'}) \wedge (\exists v, v')\left[\left(v \underset{\mathscr{M}}{\Longrightarrow} v'\right) \wedge (x \subset v')\right]\right]?$$

Question 1 is the *viral detection problem*, Question 2 constitutes the *viral evolution problem*, and Question 3 is the *viral computability problem*. Each of these three questions is a *decidability problem*, i.e. a computability problem for total functions (cf. [30]). More specifically, these total functions are mappings from the set $\mathcal{M}$ to the set $\mathcal{V}$. Additionally, a decidability problem requires an algorithm that provides a Yes/No-answer in finite time. Cohen gave a negative answer to the first two problems by reduction to the halting problem (see [13] for the halting problem and for reducibility). With respect to the viral computability problem, the answer is positive: Any string that can be computed by the universal Turing machine $\mathscr{U}$ may be a viral evolution. Cohen draws the conclusion that viral evolution exhibits the same computational abilities and power of a Turing machine, so that we have a bijection from $\mathcal{M}$ onto $\mathcal{V}$; in particular, there is also a universal viral machine $\mathscr{V}$. If we consider this result as a theorem, then we have the following result associated with it:

**Corollary 23.** *There are exactly $\omega$ viruses.*

## 3.2 Code Camouflage Techniques

*Code camouflage* is here the technical term for the code-writing techniques employed by virus designers that aim at evading detection and analysis by AVT.[12] From garbage code insertion to real code mutation s.t. for a virus $v$ s.t. $v \underset{\mathscr{M}}{\Longrightarrow} v'$ where $v'$ is highly dissimilar w.r.t. $v$, anything goes; the objective is to disable the detection and analysis techniques employed by the available AVT. There are three main code camouflage techniques, to wit, encryption, polymorphism, and metamorphism (e.g., [31]).

In order to define each of these camouflage techniques we now need a more practical definition of a virus that changes its code in its self-replications:

**Definition 24.** A pair $(v, v') \in (\mathscr{M}, V)$ s.t. $v \underset{\mathscr{M}}{\Longrightarrow} v'$ where $v' \neq v$ but $v, v' \in \mathcal{L}(\mathscr{M})$ is a *mutating virus*. Given a mutating virus $(v, v')$, we call any replication or copy $v'$ a *mutation* or *generation* of *root virus v*.

**Proposition 25.** *Let $v, v' \in \mathcal{L}(\mathscr{M})$ where $(v, v')$ is a mutating virus. Then, the set $\mathcal{L}(\mathscr{M})$ is countable.*

The proof is intuitively obvious; we let $v_0$ be the root virus and $v_i$, $i = 1, 2, ...$, be its successive mutations or generations.

Let us begin by considering that a mutating virus $(v, v')$ has two parts to it, to wit, a *main body* and a *decryption code*, or *decryptor* for short. The three main code camouflage techniques can be briefly defined as follows:

**Definition 26.** Let $v, v' \in \mathcal{L}(\mathscr{M})$ for some given Turing machine $\mathscr{M}$ s.t. $v \underset{\mathscr{M}}{\Longrightarrow} v'$. If for all $v' = v_{i=1,2,...}$ s.t.

1. the main body of $v_i$ is encrypted and from generation to generation the decryptor remains constant, we say that the pair $(v, v')$ is an *encrypted virus*.

2. the main body of $v_i$ remains constant, albeit encrypted, and the decryptor changes from generation to generation, we say that the pair $(v, v')$ is a *polymorphic virus*.

3. both the main body of $v_i$ and the decryptor change from generation to generation, we say that the pair $(v, v')$ is a *metamorphic virus*.

[12]To be more precise, we speak of code camouflage w.r.t. viral detection and of *code obfuscation* w.r.t. viral analysis. In any case, we can also employ the term *obfuscation* for both detection and analysis.

From this definition, we can make a further distinction by claiming that only polymorphic and metamorphic viruses are truly *evolving* viruses, with non-evolving viruses constituting the class of *monomorphic* viruses, or viruses with a single form. Encryption is considered essentially a *static* camouflage technique; in effect, encryption merely *hides* the main body of a virus, reason why a common antivirus software can easily detect an encrypted virus, as will be seen in the next Section. This leaves us with two *dynamic* camouflage techniques of interest for this study, to wit, polymorphism and metamorphism.

## 3.3 Antiviral Techniques vs. Code Camouflage

Let us divide the antiviral techniques into two major classes, to wit, also *static* and *dynamic* techniques. Although static analysis may be quite advanced (e.g., heuristic and spectral analysis) it essentially tries to find a *viral signature*, i.e. a string of typically sixteen unique bytes that remains unchanged in all the self-copies of a given virus, by scanning or searching in the code. This is a highly successful, usually no-false positives antiviral technique that keeps AVT in the market despite the fast proliferation of viruses in the wild: Every new signature that is identified is added to the software's database. This antiviral technique is adequate for the detection of both simple or encrypted viruses; in this latter case, the code pattern of the decryptor is all that the AVT needs for detection.

A crucial point for the anti-antiviral success of metamorphic viruses lies in the fact that the code mutations in each replication leave fewer than the unique sixteen bytes required for detection. For viruses whose code mutates when replicating themselves static analysis is insufficient and antiviral software resorts to dynamic techniques; instead of a signature database, these techniques compare the suspicious behavior of a given code with a database of known viral behavior. Code emulation, a technique that emulates the decryption process of a given evolving virus in a sandboxed environment, is the most advanced dynamic technique.[13]

Although polymorphic viruses generally pose a challenge to AVT they can be detected by employing emulation. This is not the case with metamorphic viruses, because, as a matter of fact, against what I stated above in Definition 26.3, they need not even have a constant body, let alone a decryptor: All they need is a *metamorphic engine*. In particular, in a metamorphic virus data is carried as code. I next compare these two types of virus from the viewpoint of the integration of data and program in the viral code. Again, my starting point is a core computability result, to wit, the *s-m-n theorem*.

## 3.4 Data and Programs: The *s-m-n* Theorem

The first step is the formal elaboration on the distinction between *data* and *programs*. Let us consider a universal Turing machine $\mathscr{U}$ s.t. given any partial computable function $\varphi_e(x)$, where $e$ is the index of a given Turing machine $\mathscr{M}$ and $x$ denotes some data, we have $\mathscr{U}(e,x) = \varphi_e(x)$. Recall that a Turing machine $\mathscr{M}_e$ just is the program $\mathscr{P}_e$; doing justice to its name, the universal Turing machine $\mathscr{U}$ is the unique Turing machine that, given any index $e$ and data $x$, computes exactly the same output as $\varphi_e(x)$. Clearly, if $\varphi_e^m = \varphi_e(x_1, ..., x_m)$ is the $m$-ary partial computable function computed by the Turing machine with index $e$, then we have

$$(\maltese) \qquad \psi(e, x_1, ..., x_m) = \varphi_e(x_1, ..., x_m)$$

[13] See [3] and Ször [32] for comprehensive discussions of malware detection techniques.

where the left side of (✠) corresponds to the function computed by the universal Turing machine $\mathscr{U}$ and the right side of (✠) corresponds to the function computed by the Turing machine $\mathscr{M}_e$ on the input $x_1, ..., x_m$. The meaning of this equation is that given a universal Turing machine $\mathscr{U}$ we are no longer allowed to distinguish between data and programs; in the words of virology, both data and programs are potential virus hosts.

Consider now the following result first formulated in [33], the same source as for Kleene's fixpoint theorem:

**Theorem 27.** *(s-m-n theorem) Given $m, n \in \mathbb{N}$, consider the computable function of $m + n$ variables:*

$$\psi = \varphi\left(y_1, ..., y_m, x_1, ..., x_n\right)$$

*Let now $e \in \mathbb{N}$ define $\psi$ computably. Then, there is a primitive computable function $\mathcal{S}_n^m\left(z, y_1, ..., y_m\right)$ s.t. for fixed numbers $c_{y_1}, ..., c_{y_m}$*

$$(*) \qquad \mathcal{S}_n^m\left(e, c_{y_1}, ..., c_{y_m}\right)$$

*computably defines*

$$\psi' = \varphi\left(c_{y_1}, ..., c_{y_m}, x_1, ..., x_n\right)$$

*as a function of the remaining variables.*

Clearly, then we have

$$\psi \underset{c}{=} \psi' \qquad \Leftrightarrow \qquad (*)$$

it being the case that the $c_{y_1}, ..., c_{y_m}$, each $c_{y_i}$ denoting the substitution of variable $y_i$ by some constant $c$ s.t. $c_{y_i} = k_i$ for $i = 1, ..., m$, correspond to data. The juice to be extracted from this at first sight innocuous theorem is that data can be incorporated in a program as a sub-program or, what is the same formally:

$$\varphi_{\mathcal{S}_n^m(e, y_1, ..., y_m)}\left(x_1, ..., x_n\right) = \varphi_e\left(y_1, ..., y_m, x_1, ..., x_n\right)$$

## 3.5 Polymorphic vs. Metamorphic Viruses

It will now be useful, with [34] and [35], to consider an evolving virus from the viewpoint of its core constituent:

**Definition 28.** The *kernel* of a virus is a 4-tuple

$$\mathcal{K} = \left(I\left(d, p\right), T\left(d, p\right), D\left(d, p\right), S\left(p\right)\right)$$

where $d$ and $p$ denote respectively *data* and *program*, $I\left(d, p\right)$ and $T\left(d, p\right)$ are two computable predicates, and $D\left(d, p\right)$ and $S\left(p\right)$ are two computable functions.

In practical terms, $I\left(d, p\right)$ is the *infection trigger condition*, $D\left(d, p\right)$ is the *payload routine*, $T\left(d, p\right)$ is the *payload trigger condition*, and $S\left(p\right)$ is a *selection function*. Note here that, in contrast to the remaining constituents of a kernel, function $S$ acts solely on programs, altogether neglecting data. We denote that conditions $T\left(d, p\right)$ and $I\left(d, p\right)$ are satisfied by writing respectively $T\left(d, p\right)$ ♪ and $I\left(d, p\right)$ ♪.

**Definition 29.** Let $v, v'$ be two total computable functions. The pair $(v, v') \in \mathcal{L}(\mathscr{M})$ for some Turing machine $\mathscr{M}$ is a *polymorphic virus with two forms* if, for every $x$, it satisfies

$$(\bigstar) \qquad \varphi_{v(x)}(d,p) = \begin{cases} D(d,p) & \text{if } T(d,p) ♪ \\ \varphi_x(d, p(v'(S(p)))) & \text{if } I(d,p) ♪ \\ \varphi_x(d,p) & \text{otherwise} \end{cases}$$

and

$$(\bigstar') \qquad \varphi_{v'(x)}(d,p) = \begin{cases} D(d,p) & \text{if } T(d,p) ♪ \\ \varphi_x(d, p(v(S(p)))) & \text{if } I(d,p) ♪ \\ \varphi_x(d,p) & \text{otherwise} \end{cases}$$

Intuitively put, in ★ function $\varphi_{v(x)}(d,p)$ outputs (1) the payload $D(d,p)$ if the payload trigger condition $T(d,p)$ is satisfied, (2) an evolved replica if the infection trigger condition $I(d,p)$ is satisfied (i.e. the prospective host has not already been infected), or (3) itself otherwise. (Note here Kleene's fixpoint theorem in 3, a behavior that can be called *imitation*.) In ★′ the roles of functions $v$ and $v'$ are inverted and we have a polymorphic virus with only two forms. This is to be contrasted with the following definition, which gives us the infinite set $V^{\omega}_{Poly} = \{v(m,x) \mid m \in \mathbb{N}\}$ of polymorphic generations:

**Definition 30.** A total computable function $v(m,x)$ is a *polymorphic virus with an infinite number of forms* if, for all $x$, it satisfies

$$\varphi_{v(m,x)}(d,p) = \begin{cases} D(d,p) & \text{if } T(d,p) ♪ \\ \varphi_x(d, p(v(m+1, S(p)))) & \text{if } I(d,p) ♪ \\ \varphi_x(d,p) & \text{otherwise} \end{cases}$$

and if, for every $m \neq n$, we have $v(m,x) \neq v(n,x)$.

Clearly, we have $V^{\omega}_{Poly} = \mathcal{L}_P(\mathscr{M})$ where $\mathcal{L}_P(\mathscr{M})$ specifies a polymorphic largest viral set for some Turing machine $\mathscr{M}$. Figure 1 displays the general structure of a polymorphic engine.

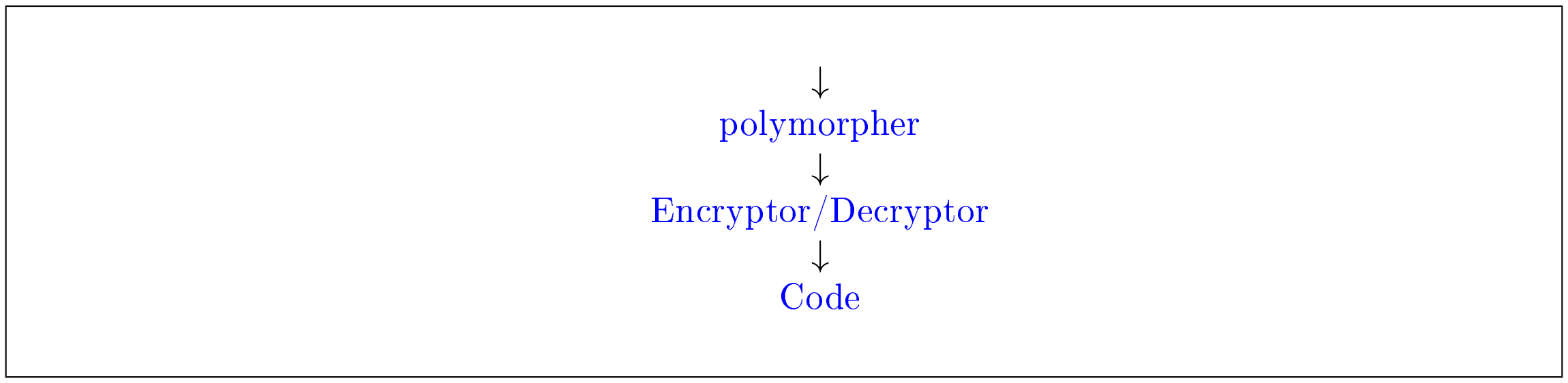

Figure 1: General structure of a polymorphic engine.

As for the metamorphic viruses, we have the following results:

**Definition 31.** Let $v, v'$ be two different total computable functions. The pair $(v, v')$ is a metamorphic virus if, for all $x$, it satisfies

$$\varphi_{v(x)}(d,p) = \begin{cases} D(d,p) & \text{if } T(d,p)\,\text{♪} \\ \varphi_x(d, p(v'(S(p)))) & \text{if } I(d,p)\,\text{♪} \\ \varphi_x(d,p) & \text{otherwise} \end{cases}$$

and

$$\varphi_{v'(x)}(d,p) = \begin{cases} D'(d,p) & \text{if } T'(d,p)\,\text{♪} \\ \varphi_x(d, p(v(S'(p)))) & \text{if } I'(d,p)\,\text{♪} \\ \varphi_x(d,p) & \text{otherwise} \end{cases}$$

where $T(d,p)$, $I(d,p)$, $D(d,p)$, and $S(p)$ are different from, respectively, $T'(d,p)$, $I'(d,p)$, $D'(d,p)$, and $S'(p)$.

This definition can be generalized to any tuple of total computable functions giving us the set $V^{\omega}_{Meta} = \left\{ v^m(S^n(p)) \,|\, m, n \in \{'\}^{\mathbb{N}} \text{ and } m \neq n \right\} = \mathcal{L}_M(\mathscr{M})$ where $\mathcal{L}_M(\mathscr{M})$ specifies a metamorphic largest viral set for some Turing machine $\mathscr{M}$. To be noted is the fact that the main difference between the polymorphic and the metamorphic viruses lies in the fact that $S(p) \neq S'(p)$. In practice, this formalization is realized in two very distinct mutation engines, as shown in Figures 1 and 2.

↓
1. Disassembly/Depermutation
↓
2. Compression
↓
3. Permutation
↓
4. Expansion
↓
5. Reassembly

Figure 2: General structure of a metamorphic engine.

# 4 Viral Metamorphism and Formal Grammars

## 4.1 The Viral Decision Problem and Formal Language Theory

It is assumed here that a successful virus is a largest viral set that escapes detection by an AVT. As already claimed above, metamorphic viruses are currently the most successful in the wild. Thus, it is important to elaborate on why metamorphism poses a greater challenge to AVT than polymorphism. To put it in a nutshell, we say that metamorphic viruses constitute a more intractable decision problem than polymorphic ones. In effect, broadly conceived *tractability* is a measure of *practical* *computability*, or how difficult it is to solve a given decision problem for a specific language; we say that a decision problem is intractable, or unsolvable, if there is no efficient algorithm that solves it.

It will now be relevant to remark that metamorphism is a code property that finds applications other than in malware. In particular, metamorphism is employed in the software industry to protect proprietary code from attacks (e.g., [36]). It is precisely the intractability of metamorphic coding that secures the protection of the proprietary code, so the reader can have an intuitive conception of how intractable metamorphic code is. With respect now to viral metamorphism, the decision problem for pure metamorphic viruses is unsolvable, i.e. given a virus $v$ and a largest viral set $\mathcal{L}_M(\mathscr{M})$, there is no algorithm that answers the question

$$(\mathrm{DP}_{viral}) \qquad v \overset{?}{\in} \mathcal{L}_M(\mathscr{M})$$

if $v$ is a pure metamorphic virus.

I shall shortly give a definition of a pure metamorphic virus. In order to analyze this it will be useful to move from the vocabulary and formalism of F. Cohen, which directly captures virology, to that of N. Chomsky, which deals with grammars and languages. As seen above, viruses just are computer programs, and these in turn just are strings of symbols of a given language. The viral decision problem can then be posed simply as the general decision problem for a given formal language, i.e. as

$$(\mathrm{DP}) \qquad w \overset{?}{\in} L$$

where $w$ is a given word and $L$ is a specific language. This move will then allow us to approach the intractability of metamorphic viruses in the framework of the complexity classes for languages.

Recall from above that for every virus we have $v \in V$ where $V \subseteq \Sigma^*$. Note now that $\Sigma^*$ is the set of all the strings, including the empty string $\lambda$, that can be built over alphabet $\Sigma$. Any subset $V^{(i)} \subseteq \Sigma^*$ is a *language* if $\Sigma$ contains only terminal symbols. The strings that constitute a given language are constructed according to the rules of a specific grammar. The following definition is very general, but Table 1 provides specifications. (See [13] for further contents).

**Definition 32.** A *formal grammar* is a 4-tuple $G = (V, T, S, P)$, where $(V \cup T) = \Sigma \neq \emptyset$ for the finite disjoint sets $V$ and $T$ of variable symbols and terminal symbols, respectively, is called the alphabet, $S \in V$ is the start symbol, and $P$ is a finite set of production rules $r_P$ (abbr.: $r$) of the form $\alpha \to \beta$, read "$\alpha$ is rewritten as $\beta$," where $\alpha$ is the left-hand side of the rule (abbr.: LHS), $\alpha \in (V \cup T)^+$, $|V(\alpha)| \geq 1$, and $\beta \in (V \cup T)^* = \Big((V \cup T)^+ \cup \{\lambda\}\Big)$ is the right-hand side of the rule (RHS), so that the empty symbol "$\lambda$" is allowed only on the RHS.

Note in this definition the notation $\Sigma^+$ to denote $\Sigma \setminus \{\lambda\}$. By a slight abuse of terminology, $G$ is also called a generative system and the language $L$ generated by $G$ is defined as:

**Definition 33.** A *formal language* $L$ is a set of strings of terminal symbols generated by a formal grammar $G$, i.e.

$$L(G) = \Big\{ w \in T^* \mid S \overset{*}{\underset{G}{\Longrightarrow}} w \Big\}$$

where "$w = a_1 a_2 ... a_k$" denotes a string of $k$ terminal symbols (the concatenation of the terminal symbols $a_1, a_2, ..., a_k$) called a *word*, "$\Longrightarrow$" denotes a derivation step such that we have $S \overset{n}{\underset{G}{\Longrightarrow}} \vartheta$ for the sentential form $\vartheta \in (V \cup T)^*$ in $n$ derivation steps, and "*" (Kleene star) denotes the reflexive and transitive closure of the relation $\underset{G}{\Longrightarrow} \subseteq (V \cup T)^+$. A *derivation* of a word $w \in L(G)$ from $S \in V_G$ is a finite sequence of steps each of which is the application of some production rule in $\{r_i\}_{i=1}^{k} \subseteq P_G$:

$$\mathscr{D}_w = S \xRightarrow[r_1]{1} \vartheta_1 \xRightarrow[r_i]{2} \vartheta_2 \xRightarrow[r_i]{3} ... \xRightarrow[r_i]{n} w$$

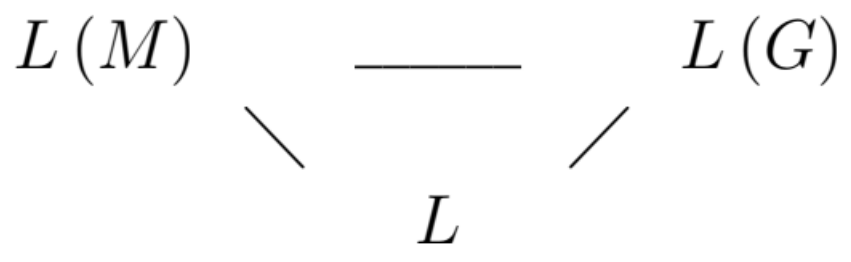

Figure 3: The basic postulate of the Chomsky hierarchy.

A derivation $\alpha \Longrightarrow \beta$ is called a *leftmost derivation*, denoted by $\alpha \Longrightarrow_l \beta$, (a *rightmost derivation*, denoted by $\alpha \Longrightarrow_r \beta$) if at each step a production rule is applied to the leftmost (respectively, rightmost) variable in $\alpha$.

The sequence of the $k$ rules applied in $n$ steps in a derivation $\mathscr{D}_w$ is called a *parse* and is defined as:

$$\mathscr{P}_w = r_{1,1} r_{i,2} r_{i,3} ... r_{i,n}$$

A parse of a word $w$ that is represented as a tree is called a parse tree for $w$, denoted by $\mathscr{T}_w$. A parse is said to be *left* (*right*) if the rules are applied in a leftmost (respectively, rightmost) sequence.

Importantly, there is a word derivation or parse iff there is a *recognizer* that accepts the given word as an element in $L(G)$. This allows for a reformulation of the definition of a language as follows:

**Definition 34.** A *formal language* $L$ is a set of words accepted by a recognizer $M$:

$$L(M) = \{w \in T^* \mid w \text{ is accepted by } M\}$$

Figure 3 shows the basic postulate of the Chomsky hierarchy, to wit, for every language $L \neq \emptyset$ there is both a grammar that generates it and a recognizer that accepts it. This postulate provides us with the following core result:

**Theorem 35.** *Given a grammar $G$ and a recognizer $M$, $M$ decides whether $L(G) = L(M)$. Thus, $M$ is a solution for the decision problem for a given language $L = L(G)$.*

The proof of this theorem depends on the recognizer and its corresponding grammar type; see [13] for the proofs of the four languages of the Chomsky hierarchy. I give a sketch of the proof for the CFGs, because I shall require this result below: If $G$ is a CFG, then for any word $w \in L(G)$ there is a pushdown automaton $M$ that constructs a leftmost derivation for it s.t. $L(G) = L(M)$. Note that the CFGs have the property that there is a one-to-one correspondence between leftmost derivations and left parses.

It will now be convenient to elaborate briefly on the simplest grammars in the Chomsky hierarchy, the linear grammars.

**Definition 36.** A *linear grammar* is a grammar $G = (V, T, S, P)$ with rules of the form $\alpha \to \beta$ in which $\alpha \in V$, $\beta \in (V \cup T)^*$, and $|V(\beta)| \leq 1$. A linear grammar can be *right-* or *left-linear* if it has rules of the form $\alpha \to vA$ or $\alpha \to Av$, respectively, for $A \in V^*, v \in T^*$. In a linear grammar, every derivation is either a leftmost or a rightmost derivation. A left- or right-linear grammar is a *regular grammar*. The words generated by a regular grammar follow a given pattern called *regular expressions*.

Table 1: The Chomsky hierarchy. (Adapted from [13])

| **Type** | **Name of Grammar** | $\alpha \to \beta$ | **Language** | **Recognizer** | **Decision Problem** |
|---|---|---|---|---|---|
| 0 | Unrestricted (UG) | $\alpha \in (V \cup T)^{+}, \lvert V(\alpha) \rvert \geq 1$<br>$\beta \in (V \cup T)^{*}$ | Recursively enumerable (REL) | Turing machine | Unsolvable |
| 1 | Context-sensitive (CSG) | $\alpha, \beta \in (V \cup T)^{+}, \lvert\beta\rvert \geq \lvert\alpha\rvert$<br>$\alpha = \gamma A \delta$<br>$\beta = \gamma X \delta$<br>$\gamma, \delta \in (V \cup T)^{*}$<br>$X \in (V \cup T)^{+}$ | Context-sensitive (CSL) | Linear-bounded automaton | **PSPACE**-complete |
| 2 | Context-free (CFG) | $\alpha \in V, \lvert\alpha\rvert = 1$<br>$\beta \in (V \cup T)^{*}$ | Context-free (CFL) | Pushdown automaton | **P** |
| 3 | Regular | $\alpha \in V, \lvert\alpha\rvert = 1$<br>$\beta = Av$ or $vA$<br>$A \in V^{*}, v \in T^{*}$ | Regular (REG) | Finite automaton | **P** |

**Example 37.** Consider the following grammar $G_a = (\{S, A\}, \{a\}, S, P)$ with

$$P = \left\{ \begin{array}{cc} (1) & S \to aS \\ (2) & S \to aA \\ (3) & A \to a \\ (4) & A \to \lambda \end{array} \right\}.$$

This is a right-linear grammar. The language generated by this grammar is defined as:

$$\begin{aligned} L(G_a) &= \left\{a^n \in T^+ | \, n \geq 1\right\} \\ &= \left\{a, a^2, a^3, ..., a^i, ...\right\} \end{aligned}$$

The regular expression for this language is $a^+$. The derivation of the word $a^4$ is given by

$$\mathscr{D}_{a^4} = S \underset{1}{\Longrightarrow_r} aS \underset{1}{\Longrightarrow_r} aaS \underset{2}{\Longrightarrow_r} aaaA \underset{3}{\Longrightarrow_r} aaaa$$

Figure 4 displays the corresponding parse tree.

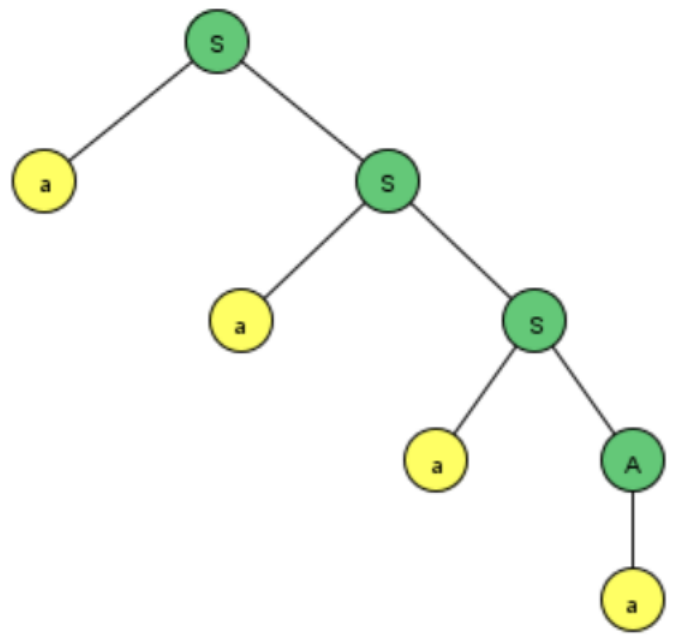

Figure 4: Parse tree for the word $a^4$ generated by grammar $G_a$.

## 4.2 Viral Polymorphism and Polymorphic Grammars

I begin this Section by giving an example of a polymorphic grammar that has starred in the literature (e.g., [14]).

**Example 38.** Consider the grammar $G_{rt} = (\{A, B\}, \{a, b, c, d, r, t\}, S, P)$ where the production set is as follows:

$$P = \left\{ \begin{array}{cc} (1) & S \to aS \,|\, bS \,|\, cS \,|\, rA \\ (2) & A \to aA \,|\, bA \,|\, cA \,|\, dA \,|\, tB \\ (3) & B \to aB \,|\, bB \,|\, cB \,|\, dB \,|\, \lambda \end{array} \right\}$$

This is a right-linear grammar that generates words of the form

$$\Theta^* r \Theta^* t \Theta^*$$

where $\Theta = \{a, b, c, d\}$. Note that this grammar has in fact fourteen production rules; we managed to reduce this number to three by collecting rules with the same LHS as alternative RHSs. This allows for enumeration of the rules as

$$(r1.1) \quad S \to aS$$

$$(r1.2) \quad S \to bS$$

etc.

The simplest word this grammar generates is $rt$, namely when we have $\lambda r \lambda t \lambda$. The word *arcccdt* belongs to the language generated by this grammar. The derivation of this word is given by:

$$\mathscr{D}_{arcccdt} = S \underset{1.1}{\Longrightarrow} aS \underset{1.4}{\Longrightarrow} arA \underset{2.3}{\Longrightarrow} arcA \underset{2.3}{\Longrightarrow} arccA \underset{2.3}{\Longrightarrow} arcccA \underset{2.4}{\Longrightarrow} arcccdA \underset{2.5}{\Longrightarrow}$$

$$arcccdtB \underset{3.5}{\Longrightarrow} arcccdt$$

See Figure 5 for the parse tree for this word.

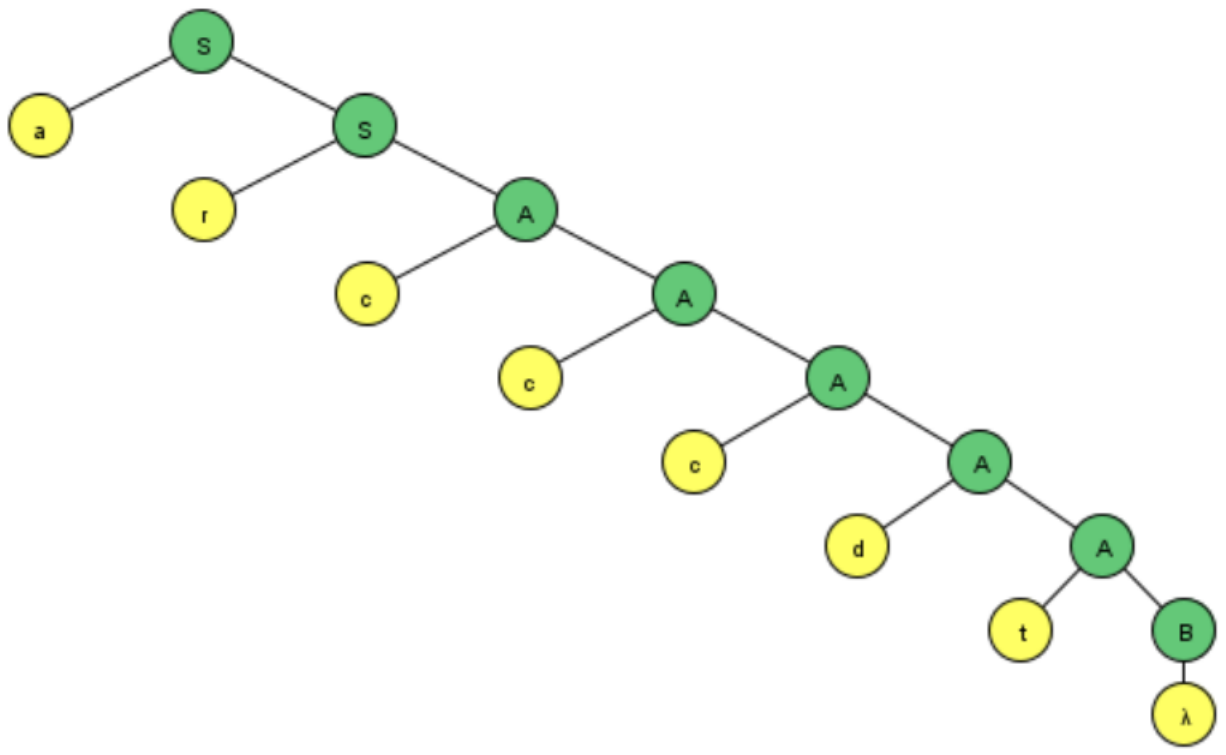

Figure 5: Parse tree for the word *arcccdt* generated by the polymorphic grammar $G_{rt}$.

**Definition 39.** With respect to Example 40, let us call the word $rt \in L(G_{rt})$ a *root word* of $L(G_{rt})$ and the expression $\Theta^* r \Theta^* t \Theta^*$ its *mutation pattern*.

Some basic mutation rules are as follows:

**Definition 40.** Let $w \overset{n}{\Longrightarrow} w', n > 0$, be a mutation of $w$.

1. If $|w'| > |w|$, mutation $w'$ is called an *expansion* of $w$; otherwise, it is called a *compression* of $w$.
2. If $|w'| = |w|$ but $w' \neq w$ while the letters in $w'$ are exactly the same as in $w$, we say that $w'$ is a *permutation* of $w$.

In Example 40, there is a single root word for the given language, but this need not be the case; a language may have as many root words as required. If the grammar at hand is a regular grammar, the mutation pattern can be given by a regular expression. It should be easy to see that a mutation pattern should be constructed in such a way as to allow for potentially infinite mutations of a given root word.

**Definition 41.** Let $G = (V, T, S, P)$ be a formal grammar. $G$ is called a *polymorphic grammar* if it generates multiple mutations of root words of $L(G)$.

**Example 42.** Let us consider again the grammar in Example 40. If the root word $rt \in L(G_{rt})$ is composed of two instructions solely, to wit, $r$ and $t$, then the components $\Theta^+$ in $\Theta^* r \Theta^* t \Theta^*$ correspond to garbage code insertion. We may make it so that $L(G_{rt})$ generates further polymorphic code by introducing production rules for the permutation of the instructions, so that we have also a root word $tr \in L(G_{rt})$ and a mutation pattern $\Theta^* t \Theta^* r \Theta^*$.

We can find further ways to make it more difficult to be able to identify words of this language, but this grammar generates a decidable language, as is shown by the fact that (unambiguous) parse trees can be built for every one of its words. (See Fig. 5.) This is a consequence of employing a regular language, for which the decision problem is in the complexity class **P** (cf. Table 1). One may in fact employ a CSG to generate a polymorphic virus, but this will only move the problem from class **P** to class **PSPACE**-complete. To be sure, this might make it far more difficult for an AVT to identify a polymorphic virus.[14] Nevertheless, if you check Table 1 for the status of the

[14]For the sake of self-containment, we provide here the essentials of the standard complexity classification. The reader is referred to Arora & Barak (2009) for a comprehensive textbook in complexity theory.

**Definition 43.** Let $\mathscr{M}$ denote a Turing machine on an input string $x \in \Sigma^*$ such that $|x| = n$. Let $T(x)$ denote the *time complexity* of $\mathscr{M}$, i.e. the maximum number of moves that $\mathscr{M}$ makes before halting on input string $x$. We say that *$\mathscr{M}$ runs in polynomial time*, or that *$\mathscr{M}$ has polynomial-time complexity*, iff there is a polynomial $p$ such that $\mathscr{M}$ runs in time bounded by $p(n) = \mathscr{O}(n^k), k \in \mathbb{N}$, such that $T(x) \leq p(n)$ for $n = 0, 1, ...$

*Space complexity* for $\mathscr{M}$, denoted by $S(n)$, is defined in the same terms w.r.t. the maximum number of tape cells scanned by $\mathscr{M}$. Recall that the notation $f(n) = \mathscr{O}(n^k)$ denotes that $\mathscr{O}(n^k)$ is an (asymptotic) upper bound for $f(n)$. Consider now that a Turing machine can be either deterministic (if the next step in a computation is unique) or non-deterministic (otherwise), and let us denote them by DTM and NTM, respectively.

**Definition 44.** Let $n \in \mathbb{N}$ be given for some input string $x \in \Sigma^*$ such that $|x| = n$.

(1) We say that a DP is solvable in *deterministic polynomial time* iff there is a DTM $\mathscr{M}$ that solves it in a number of moves bounded above by a polynomial $p(n)$. This defines the complexity class **P** that contains all languages that are decidable by a DTM with polynomial-time complexity, i.e.

$$\mathbf{P} = \bigcup_{k \in \mathbb{N}} \mathrm{DTIME}\left(n^k\right).$$

(2) We say that a DP is solvable in *nondeterministic polynomial time* iff there is an NTM $\mathscr{M}$ that solves it in a number of moves bounded above by a polynomial $p(n)$. This defines the complexity class **NP** that contains all languages that are decidable by an NTM with polynomial-time complexity, i.e.

$$\mathbf{NP} = \bigcup_{k \in \mathbb{N}} \mathrm{NTIME}\left(n^k\right).$$

Theoretically, one is told that the problems in **P** are those for whose solution there is an efficient algorithm, whereas those in **NP** can have their solution efficiently verified; in practice, problems in **P** are tractable while **NP**-problems may be intractable. The **NP**-complete class is constituted by the hardest-to-solve problems in the **NP** class.

Completeness works in the same way for the remaining complexity classes. The complexity class **PSPACE** contains all languages that are decidable by a DM with polynomial-space complexity, i.e. $\mathbf{PSPACE} = \bigcup_{k \in \mathbb{N}} \mathrm{DTIME}\left(n^k\right)$.

decision problem for the different languages you will see that only the RELs have an unsolvable decision problem; hence, this polymorphic language can pose a challenge to an AVT only if it is generated by a UG.

# 5 Pure Metamorphic Viruses and the van Wijngaarden grammars

## 5.1 The Viral Decision Problem for a Pure Metamorphic Virus

From the contents above, it should be evident that metamorphic viruses successfully evade antiviral detection and analysis by keeping their *functionality*—their meaning—while changing their *appearance*—the code—in each and every self-replication. This, however, can be attained by means of polymorphism, so there must be additionally something that accounts for the greater challenge posed to an AVT by a metamorphic virus. The intuition to be extracted from the formalization in Section 3.5 above is as follows: Polymorphism is the case when from mutation to mutation a virus changes its words; as seen above, a regular grammar can be employed to that end. In contrast, metamorphism entails not only this word change phenomenon, but also the change from mutation to mutation of the very mutation engine or, what is the same if we identify a grammar with the mutation engine, the grammar itself. This poses an additional challenge—and a substantial one at that—to the virus designer. It suffices to compare the general structure of a polymorphic engine (see Fig. 1) with that of a metamorphic one (Fig. 2) to see that the task ahead for a metamorphic virus designer is no simple one.

The ways to design a virus that has the evolving property of changing its appearance while preserving its functionality have been well explored in the antiviral literature (e.g., [4,8,14,38,39]). Essentially, the methods listed in these references include actions that make it more difficult for an AVT to identify and/or analyze a metamorphic virus; examples of such actions are instruction reordering, garbage code inclusion, code permutation, data reordering, etc. These actions can take up to 90% of the virus' code for the metamorphic engine.

One aspect that contributes to an undecidable problem in formal language theory is when we succeed in generating several levels to a single language s.t. we have $L\left(L_i\left(G\right)\right)$, where $i \in \mathbb{N}$. The decision problem for such a language becomes more complex to solve if there is more than one grammar involved, so that we have languages $L\left(L_i\left(G_j\right)\right)$, for $j \in \mathbb{N}$. As a matter of fact, it suffices to generate a language with only two levels, in the sense that the language $L\left(L\left(G\right)\right)$ is typically undecidable. Certainly with this mathematical result in mind, E. Filiol proposed the following formal-grammar definition for a metamorphic virus [14]:

**Definition 45.** Let $G_1 = (V, T, S, P)$ and $G_2 = (V', T', S', P')$ be two grammars where $T'$ is a set of formal grammars, $S'$ is the (starting) grammar $G_1$, and $P'$ is a rewriting system w.r.t. $(V' \cup T')^*$. A metamorphic virus is described by $G_2$ and every of its mutated forms is a word in $L\left(L\left(G_2\right)\right)$.

This definition is redundant in the following sense: It is not required that we have two distinct grammars $G_1, G_2$ as specified in Definition 45; it suffices that the language $L\left(L\left(G\right)\right)$ be generated by a *two-level grammar*, namely a *van Wijngaarden grammar* (abbr.: WG), in which case we write $L\left(L\left(\mathcal{G}\right)\right)$. Filiol leaves it as an open question whether Definition 45 is a particular case of a WG.

This class is suspected to be intractable, but note that it is currently understood (but not known) that $\mathbf{P} \subseteq \mathbf{NP} \subseteq \mathbf{PSPACE}$.

I do not aim here at closing the question, but simply at showing that a metamorphic virus can in principle be designed entirely by employing solely a WG.

**Definition 46.** Let $\mathcal{G}$ denote an arbitrary WG and let $v \in \mathcal{V}$. We say that $v$ is a *pure metamorphic virus* if $v' \in L(L(\mathcal{G}))$ for any mutated form $v'$ of $v$.

*Claim* 47. A pure metamorphic virus can be constructed.

In order to support this claim I need to show four things: (1) every word generated by a WG $\mathcal{G}$ belongs to the language $L(L(\mathcal{G}))$; (2) the number of mutations of a given virus can in principle be infinite, namely in the sense that there is a possibly infinite number of strict subgrammars associated with a single WG; (3) a virus generated by a WG is a metamorphic virus, i.e. it preserves functionality in syntactic change; and (4) the decision problem for languages generated by WGs is unsolvable.[15]

## 5.2 The van Wijngaarden Grammars

The following is a summary elaboration on WGs for the sake of self-containment, and the reader is referred to [19,20] and [21] for comprehensive discussions on the syntax and the semantics, respectively, of WGs.

**Definition 48.** A WG $\mathcal{G}$ is constituted by a *deep grammar* and a *surface grammar*. (A) The deep grammar is a 7-tuple

$$\mathcal{G}^D = \left( \underbrace{M, V, R_M}_{\mathcal{G}^1}, \underbrace{N, T, S, R_V}_{\mathcal{G}^2} \right)$$

where

$$\frac{\mathcal{G}^2}{\mathcal{G}^1} = \frac{\text{CFG}}{\text{CFG}} = \frac{\text{Hyper-level}}{\text{Meta-level}} = \frac{\text{Level 2}}{\text{Level 1}},$$

$M$, $V$, $N$, $T$, and $S$ are sets of *notions*, and $R_M$ and $R_V$ are sets of *rules* built over these notions. The sets of notions are as follows: $M$ is a set of *metanotions* and $V$ is a set of *protovariables* with $(M \cap V) = \emptyset$, $N = (M \cup V)^*$ is a set of *hypernotions*, $T$ is a set of *terminals*, and $S \subseteq N$ s.t. $S = \{\langle s \rangle\}$ is the *start hypernotion*. The elements in the set $(V \cup T)^*$ are called *protonotions*. Set $R_M$ contains the *metarules* of the grammar and set $R_V$ contains its *hyper-rules*. (See Table 2 for the form of these rules.)

(B) The *surface grammar* of a WG $\mathcal{G}$ is a 4-tuple

$$\mathcal{G}^S = (N_S, T, s, R_S)$$

where $N_S$ is a set of *strict notions*, $T$ is a set of terminal protonotions, $s \in N_S$ is the *start strict notion* derived from $\langle s \rangle \in R_V$, and $R_S$ is a set of *strict production rules*. (See Table 2 for the form of these rules.)

[15] The papers [40,41] explore the design of metamorphic viruses employing WGs in practice. The reader interested in verifying how viral code can be made to mutate under the rules of a WG is referred to these works. Our aim in the present study is strictly theoretical: We are interested in the general properties of WGs that allow for a large (possibly infinite) number of code changes to preserve functionality and at the same time to constitute an undecidable language. This can be applied to viral code for virus development or to proprietary code for intellectual property protection, other employments not being excluded.

Table 2: Grammars and levels of a WG $\mathcal{G}$ with respective rules.

| Grammar & Level | Rule | LHS | $\rightarrow$ | RHS |
|---|---|---|---|---|
| $\mathcal{G}^{D.1}$ | $r_M \in R_M$ | Metanotion $W \in M$ | :: | Meta-alternatives in $(M \cup V)^*$ |
| $\mathcal{G}^{D.2}$ | $r_V \in R_V$ | Hypernotion $\langle X \rangle \in N$ | : | Hyper-alternatives in $(N \cup T \cup \{,\})^*$ |
| $\mathcal{G}^{S}$ | $r_S \in R_S$ | Strict Notion $x' \in N_S$ | : | Strict Alternatives in $(V \cup T \cup \{,\})^*$ |

*Remark* 49. Notions are sequences of symbols $\overline{\sigma} = \sigma_0\sigma_1...\sigma_k$ s.t. $k \geq 0$ called *syntactic marks* that can be uppercase or lowercase, called respectively *large syntactic marks* and *small syntactic marks.* This distinction is crucial to segregate the different kinds of notions: Metanotions are written with large syntactic marks, protonotions are written with small syntactic marks, and hyper-notions are written with both. Terminal symbols must be followed by "symbol." For example, a symbol corresponds to the terminal symbol *a*. I remark that metanotions can also be terminal (e.g., ALPHA symbol).

The rules in $R_M$ *specify* which replacements by protonotions are admissible in $R_V$ for the metanotions and the rules in $R_V$ *describe* the properties that the words generated by $\mathcal{G}$ satisfy. Grammars $\mathcal{G}^1$ and $\mathcal{G}^2$ interact in the following way:

**Definition 50.** Let $r_M$ and $r_V$ denote respectively a metarule and a hyper-rule. A *uniform replacement rule*, written $URR_{r_M i}^{r_V j}(W,x)$, instructs that the metanotion $W$ in metarule $r_M i = W :: x$ be replaced uniformly in hyper-rule $r_V j$ by the protonotion $x$.

I abbreviate $URR_{r_M i}^{r_V j}(W,x)$ as $URR_i^j(W,x)$ and I write $URR_i^j(\mathcal{G})$ to denote a $URR_i^j(W,x)$ in $\mathcal{G}$. The interaction of the deep and surface grammars of a WG $\mathcal{G}$ is as follows: A hyper-rule $r_V j$ *derives* a strict production rule $r_S l$ whenever one or more rules $URR_i^j(W,x)$ are applied to it. Note thus that whereas the grammar levels describe the URRs of a given language the interaction between the deep and the surface grammar actually generates the language. We have the following definitions:

**Definition 51.** Given a WG $\mathcal{G}$, let

$$\widehat{\mathcal{G}} = \bigcup_{i=1}^{n} \bigcup_{j=1}^{m} URR_i^j(\mathcal{G})$$

for arbitrary $i, j \in \mathbb{N}$. Then, $\mathcal{G}^D$ *describes* the language of $\mathcal{G}$ as

$$L_{descr}(\mathcal{G}) = L_{\langle s \rangle}(\mathcal{G}) = \left\{ w \in T^* | \, \langle s \rangle \overset{*}{\underset{\widehat{\mathcal{G}}}{\Longrightarrow}} w \right\} = L\left(\mathcal{G}^D\right)$$

where $\langle s \rangle \in S$ is the start hyper-notion s.t. we have equivalently the language

$$L(\mathcal{G}) = \{w \in T^* | \, w \text{ satisfies certain properties}\}.$$

Table 3: Grammars and levels of a WG $\mathcal{G}$ and corresponding rules.

| Grammar | Subgrammar | Level | Rules |
|---|---|---|---|
| Deep ($\mathcal{G}^D$) | Metagrammar | $\mathcal{G}^1$ | Metarules |
| | Hypergrammar | $\mathcal{G}^2$ | Hyper-rules |
| Surface ($\mathcal{G}^S$) | Strict Grammars $\mathcal{G}_{x_i'}$ | | Strict Production rules |

The language *generated* by $\mathcal{G}$ is defined as

$$L_{gen}(\mathcal{G}) = L_{\mathsf{s}}(\mathcal{G}) = \left\{ w \in T^* \middle| \, \mathsf{s} \overset{*}{\underset{\widehat{\mathcal{G}}}{\Longrightarrow}} w \right\} = L\left(\mathcal{G}^S\right)$$

where $\mathsf{s} \in N_S$ is the start strict notion.

Note that $\langle s \rangle \overset{*}{\underset{\widehat{\mathcal{G}}}{\Longrightarrow}} w$ is *not* a word derivation proper: It is rather the *description* of the *-step derivation from $\langle s \rangle$ to $w$ based on $\widehat{\mathcal{G}}$. In effect, $\widehat{\mathcal{G}}$ describes the derivation, or generation, of any specific word $|w| = k$ for $k = 0, 1, 2, ...$ in the sense that a $URR_i^j(W, x)$ applied uniformly to all the hypernotions in $r_V j$ containing the metanotion $W$ *derives* a strict production rule $r_S l$ s.t. the protonotion $x$ is in $r_S l$. We thus have the set of production rules

$$R_S = \bigcup_{i=1}^{k} \widetilde{r_V i} = \bigcup_{j=1}^{\omega} r_S j$$

where $\widetilde{r_V i}$ denotes the derivation of a rule in $R_S$ from the $i$-th rule in $R_V$. A rule in $R_V$ can derive an infinite number of strict production rules, reason why parse trees for languages generated by WGs are exclusively based on the hyper-rules. This is an important remark, because a WG whose surface grammar has only a finite number of strict production rules is no different from a CFG. On the other hand, a hyper-rule need not derive any strict production rule.

As a matter of fact, rather than generating a single language the grammar $\mathcal{G}^S$ actually generates a (possibly infinite) number of strict languages. When a protonotion $x$ occupies the LHS of a strict production rule it is called a *strict notion* and is abstractly denoted by $x'$.

**Definition 52.** Let $x' \in N_S$ be a strict notion in a production rule $r_{S_{x'}}$ and such that $\mathcal{G}_{x'}$ is the strict grammar for $x'$.[16] A string of terminal symbols derived from $x'$, denoted by $w_{x'}$, such that we have $x' \overset{*}{\underset{\mathcal{G}^S}{\Longrightarrow}} w_{x'}$, is called a *strict word.* The set of all the strict words that can be generated by $r_{S_{x'}}$ is called the *strict language of* $x'$, and it is defined as:

$$L_{x'}\left(\mathcal{G}^S\right) = \left\{ w_{x'} \in T^* \middle| \, x' \overset{*}{\underset{\mathcal{G}_{x'}}{\Longrightarrow}} w_{x'} \right\}$$

Table 3 gives the complete structure of a WG $\mathcal{G}$.

I now give an example of a simple WG that captures the contents above.

[16] Note that the strict notion $x'$ can coincide with the strict start notion $\mathsf{s}$.

**Example 53.** The WG $\mathcal{G}_{Alpha}$, abbreviated as *Alpha*, is given by the following rules in its deep grammar:

$\mathcal{G}^{D}_{Alpha}$

| | | | |
|---|---|---|---|
| $(r_M1)$ | ALPHA | :: | a; b; c; ...; y; z. |
| $(r_V1)$ | s | : | ALPHA. |
| $(r_V2)$ | ALPHA | : | letter ALPHA symbol; |
| | | | ALPHA, letter ALPHA symbol. |

Metarule $r_M1$ has twenty-six meta-alternatives, hyper-rule $r_V1$ has a single hyper-alternative, and hyper-rule $r_V2$ has two hyper-alternatives. These rules describe the language

$$L\left(\mathcal{G}^{D}_{Alpha}\right) = \left\{t^n \in T^+ |\, n \geq 1, t = a, b, ..., \text{ or } z\right\}.$$

For instance, by applying $URR^{1,2}_{1}(\mathsf{ALPHA}, \mathsf{a})$, the (rightmost) derivation of the word $aaa$ is as follows:

$$\mathsf{s} \underset{r_V1}{\Longrightarrow} \mathsf{a} \underset{r_V2}{\Longrightarrow} \mathsf{a}, \underbrace{a}_{\text{letter a symbol}} \underset{r_V2}{\Longrightarrow} \mathsf{a}, \underbrace{a}_{\text{letter a symbol}} a \underset{r_V2}{\Longrightarrow} \underbrace{a}_{\text{letter a symbol}} aa$$

Note in this word derivation that the comma is "absorbed" when its role of separating strings of terminals from strings of variables is finished, and the same happens to the notion symbol when letter a symbol is substituted for the terminal $a$.[17] Grammar *Alpha* generates twenty-six strict languages, one for each of the letters of the Roman alphabet. In effect, we have:

$$L\left(\mathcal{G}^{S}_{Alpha}\right) = \bigcup_{x'=a,b,...}^{z} L_{x'}$$

We thus have the strict languages:

$$L_{\mathsf{a}} = \{a^n |\, n > 0\}$$
$$L_{\mathsf{b}} = \{b^n |\, n > 0\}$$
$$\vdots$$
$$L_{\mathsf{z}} = \{z^n |\, n > 0\}$$

The $URR^{2}_{1}(\mathsf{ALPHA}, \mathsf{a})$ applied to the entire set $R_V$ of *Alpha* gives a strict subgrammar, namely the set of strict productions $R_{S_{\mathsf{a}}}$ that generates the surface strict language $L_{\mathsf{a}}\left(\mathcal{G}^{S}_{Alpha}\right) = \{a^n |\, n > 0\}$,

$$URR(\mathsf{ALPHA}, \mathsf{a}) = R_{S_{\mathsf{a}}} = \left\{ \begin{array}{l} \mathsf{s} \,:\, \mathsf{a}. \\ \mathsf{a} \,:\, a;\, \mathsf{a}a. \end{array} \right\}$$

where I abbreviate "letter a symbol" as "$a$".Thus, the language generated by the WG *Alpha* is the union of all the strict languages generated by its strict subgrammars, i.e.

$$L\left(\mathcal{G}^{S}_{Alpha}\right) = L_{\mathsf{a}} \cup L_{\mathsf{b}} \cup ... \cup L_{\mathsf{z}}.$$

[17] Such notions are called *paranotions*.

### 5.3 WGs and Pure Metamorphic Viruses

I now retake Claim 47 by addressing the points (1) through (4) above.

(1) THE LANGUAGE $L(L(\mathcal{G}))$

From Definition 51, we can infer the identity

$$L\left(\mathcal{G}^S\right) = L\left(L\left(\mathcal{G}^D\right)\right).$$

**Theorem 54.** *Given a WG $\mathcal{G}$, every word in $L\left(\mathcal{G}^S\right)$ is a word in $L\left(L\left(\mathcal{G}^D\right)\right)$.*

*Proof.* (Informal) The deep grammar $\mathcal{G}^D$ does not generate words; rather, it specifies the URRs admissible in the grammar. The words generated in $\mathcal{G}^S$ by the rules derived from $R_V$ are those strings in which these admissible URRs are implemented. □

In effect, for every word $w \in L\left(\mathcal{G}^S\right)$, the parse tree $\mathscr{T}_w$ is built over the grammar $\mathcal{G}^D$. This is so because the number of rules in $R_S$ can be infinite, whereas the number of rules in $R_M$ and $R_V$, which correspond to CFGs, is necessarily finite.

(2) WGS GENERATE A LARGE, POSSIBLY INFINITE, NUMBER OF GRAMMARS

I stated above that what distinguishes a metamorphic from a polymorphic virus in practice is that the former changes not only the words from mutation to mutation, but also the grammar. It should now be easy to see how a WG $\mathcal{G}$ can generate a large, possibly infinite, number of grammars by its surface grammar. I speak here, of course, of the strict subgrammars. I give an example by way of a constructive proof.

**Example 55.** Consider the following deep grammar:[18]

$\mathcal{G}^D_{AlphaDig}$

| | | | |
|---|---|---|---|
| $(r_M1)$ | SEQUENCE | :: | statement; digit; letter. |
| $(r_M2)$ | ALPHA | :: | a; b; ...; z. |
| $(r_M3)$ | NUMBER | :: | zero; one; ...; nine. |
| $(r_M4)$ | N1 | :: | ALPHA$^n$ |
| $(r_M5)$ | N2 | :: | NUMBER$^m$ |
| $(r_V1)$ | program | : | statement sequence. |
| $(r_V2)$ | statement | : | N1 letter sequence, becomes symbol, N2 digit sequence. |
| $(r_V3)$ | letter | : | letter ALPHA symbol. |
| $(r_V4)$ | digit | : | digit NUMBER symbol. |
| $(r_V5)$ | SEQUENCE sequence | : | SEQUENCE; SEQUENCE sequence, SEQUENCE. |

[18] The RHS of metarule $r_M4$ ($r_M5$) instructs that the metanotion N1 (respectively, N2) is to be replaced by a sequence of $n$ letters ($m$ digits, respectively). I simplify here the rules for exponentiation; see [19-21] for comprehensive discussions.

It is easy to see that this deep grammar describes a language of words of the form "$u := v$" where $u$ is a sequence of $n$ letters, := is the becomes symbol, and $v$ is a sequence of $m$ digits. For this reason I call this grammar *AlphaDigit* (abbreviated as *AlphaDig*). For instance, the word "$age := 55$" belongs to the language described by $\mathcal{G}^D_{AlphaDig}$, i.e. the language $L\left(\mathcal{G}^D_{AlphaDig}\right)$. The representation table for this grammar is given in Table 4 in an abbreviated way; note in particular the typographical characters for the terminal notation for digits.

Table 4: Representation table (abbreviated) for $\mathcal{G}_{AlphaDig}$.

| **Terminal Notation** | **Typographical Character** |
|---|---|
| becomes symbol | $:=$ |
| letter a symbol | $a$ |
| ⋮ | ⋮ |
| letter z symbol | $z$ |
| digit zero symbol | 0 |
| ⋮ | ⋮ |
| digit nine symbol | 9 |

The start hypernotion $\langle s \rangle$ is program. The language of $\mathcal{G}^D_{AlphaDig}$ is

$$L\left(\mathcal{G}^D_{AlphaDig}\right) =$$

$$\left\{w \in T^* \,|\, w = (u := v)\,, u \in \{a, b, ..., z\}^+\,, v \in \{0, ..., 9\}^+\right\}$$

which can be further specified as:

$$L\left(\mathcal{G}^D_{AlphaDig}\right) =$$

$$\left\{\mathsf{u} := \mathsf{v} \,|\, \mathsf{u} \in \left\{URR\left(\mathsf{ALPHA}, \mathsf{x}_i\right)_{i=\mathsf{a}}^{\mathsf{z}}\right\}^+, \mathsf{v} \in \left\{URR\left(\mathsf{NUMBER}, \mathsf{y}_j\right)_{j=0}^{9}\right\}^+\right\}$$

$L\left(\mathcal{G}^D_{AlphaDig}\right)$ contains already all the words of $L\left(\mathcal{G}^S_{AlphaDig}\right)$ as non-terminal words, even if the syntactic mark ":=" occurs in $\mathcal{G}^D_{AlphaDig}$ as a terminal.

The following are the production rules of the WG $\mathcal{G}^S_{AlphaDig}$:

$$\mathcal{G}^S_{AlphaDig}$$

| | | | |
|---|---|---|---|
| $(r_S1)$ | program | : | statement sequence. |
| $(r_S2)$ | statement sequence | : | statement; statement sequence, statement. |
| $(r_S3)$ | statement | : | n letter sequence, becomes symbol, m digit sequence. |
| $(r_S4)$ | n letter sequence | : | letter; letter sequence, letter. |
| $(r_S5)$ | letter | : | letter a symbol; ...; letter z symbol. |
| $(r_S6)$ | m digit sequence | : | digit; digit sequence, digit. |
| $(r_S7)$ | digit | : | digit zero symbol; ...; digit nine symbol. |

Note that this set of production rules is infinite, because for each program we must specify beforehand the length $n$ of the letter sequence (rule $r_S4$) and the length $m$ of the digit sequence (rule $r_S6$). That is, each program is specified as an $n+m$ program, for $n, m \in \mathbb{N}$. For instance, the word "$a := 0$" is generated by the strict grammar $\mathcal{G}^{S[1+1]}_{AlphaDig}$. The language generated by this WG is thus the infinite (but recursively enumerable) strict language:

$$L_{\mathsf{program}}\left(\mathcal{G}^{S}_{AlphaDig}\right) = \{a := 0, a := 1, ..., age := 55, ...\}$$

(3) WGS NATURALLY PRESERVE MEANING IN SYNTACTIC CHANGE

Above, I defined metamorphism generally as the code property that preserves meaning in syntactic change. WGs do this by specifying a fixed set of Cartesian products $Type \times Value$ in what can be called an *intrinsic semantics* (cf. [21]).

**Definition 56.** Let an atomic metarule be semantically defined as

$$(r_M) \qquad \underbrace{W}_{\text{Type}} :: \underbrace{v}_{\text{Value}}$$

where $W$ denotes a metanotion and $v$ denotes a protovariable. A WG semantics $\mathfrak{S}$ is the (possibly infinite) set of pairs $(W, v)$ formed by a pairing function $\mathfrak{s} : M \longrightarrow V$ that pairs types with values as stipulated in the metarules. I accordingly call this a *type-by-value uniform replacement semantics* (abbr.: TVURS).

It is the coincidence $\mathfrak{S}_{L_\mathsf{s}(\mathcal{G}^S)} = \widehat{\mathcal{G}}$, where $\mathfrak{S}_{L_\mathsf{s}(\mathcal{G}^S)}$ denotes the semantics of a strict language generated by a WG $\mathcal{G}$, that provides a WG language with metamorphism. See [21] for the formal aspects of this coincidence.

**Example 57.** Let us consider again the WG *AlphaDig* (for the sake of simplicity, I omit the URRs for exponentiation). We have the set

$$\widehat{\mathcal{G}_{AlphaDig}} = \left\{ \begin{array}{c} (\mathsf{SEQUENCE}, \mathsf{statement}), (\mathsf{SEQUENCE}, \mathsf{letter}), (\mathsf{SEQUENCE}, \mathsf{digit}) \\ (\mathsf{ALPHA}, \mathsf{a}), (\mathsf{ALPHA}, \mathsf{b}), ..., (\mathsf{ALPHA}, \mathsf{z}) \\ (\mathsf{NUMBER}, \mathsf{zero}), (\mathsf{NUMBER}, \mathsf{one}), ..., (\mathsf{NUMBER}, \mathsf{nine}) \end{array} \right\}$$

that specifies all the URRs for this grammar. It is easy to see that $\mathfrak{S}_{L_\mathsf{s}\left(\mathcal{G}^S_{AlphaDig}\right)} = \widehat{\mathcal{G}_{AlphaDig}}$.

(4) WGS ARE RELS

Theorems 58 and 59 below handle the fourth aspect listed above.

**Theorem 58.** *A language $L$ is a REL iff there exists a WG $\mathcal{G}$ s.t. $L = L(\mathcal{G})$.*

See [42] for the original proof. However, I remark that a language $L$ generated by a WG $\mathcal{G}$ is a REL only if the set $R_S$ in $\mathcal{G}^S$ is infinite; otherwise, it is equivalent to a CFL. Additionally, there are syntactic aspects that determine the unsolvability of a language generated by a given WG (see [20]). In particular, we have the following result:

**Theorem 59.** *It is undecidable whether or not an arbitrary WG $\mathcal{G}$ has the property that for each leftmost derivation there is at most one left parse.*

Instead of the proof, which the reader can find in [43], I give an example that shows how easy it is to construct a WG for which the decision problem is unsolvable via the unsolvability of the parsing problem.

**Example 60.** Consider the following deep grammar for a given WG $\mathcal{G}$ in which $n \geq 1$:

| | | | |
|---|---|---|---|
| $(r_M 1)$ | A | :: | aA; a. |
| $(r_M 2)$ | B | :: | A. |
| | | | |
| $(r_V 1)$ | s | : | A; B. |
| $(r_V 2)$ | A | : | $\mathsf{a}^n$. |
| $(r_V 3)$ | B | : | $\mathsf{a}^n$. |
| $(r_V 4)$ | $\mathsf{a}^n$ | : | a symbol. |

Obviously, by hyper-rule 4 this WG has a single strict language such that $L_{\mathsf{a}} = \{a\}$. Then, there are infinitely many leftmost derivations for the single left parse $\mathscr{P}_a = r_V 1, r_V 2, r_V 4$:

$$\mathsf{s} \underset{r_V 1}{\Longrightarrow}_l \mathsf{A} \underset{r_V 2}{\Longrightarrow}_l \mathsf{a} \underset{r_V 4}{\Longrightarrow}_l a$$

$$\mathsf{s} \underset{r_V 1}{\Longrightarrow}_l \mathsf{A} \underset{r_V 2}{\Longrightarrow}_l \mathsf{a}^2 \underset{r_V 4}{\Longrightarrow}_l a$$

$$\vdots$$

$$\mathsf{s} \underset{r_V 1}{\Longrightarrow}_l \mathsf{A} \underset{r_V 2}{\Longrightarrow}_l \mathsf{a}^i \underset{r_V 4}{\Longrightarrow}_l a$$

$$\vdots$$

Conversely, there are two left parses for the leftmost derivation $\mathscr{D}_a = \mathsf{s} \overset{2}{\Longrightarrow}_l \mathsf{a} \overset{i}{\Longrightarrow}_l \mathsf{a}^i \Longrightarrow_l a$, to wit:

$$(\mathscr{P}_a)_1 = r_V 1, r_V 2, r_V 4$$

$$(\mathscr{P}_a)_2 = r_V 1, r_V 3, r_V 4$$

Recall from the remarks on Problem 22 that the viral detection problem and the viral evolution problem were given negative answers by F. Cohen, but he answered the viral computability problem in an affirmative way, in the specific sense that there is a bijection from the set $\mathcal{V}$ of all viruses to the set $\mathcal{M}$ of all the Turing machines. I now state the challenge posed by a pure metamorphic virus to an AVT:

**Problem 61.** Let $(\mathscr{M}, V) \in \mathcal{V}$ be a pure metamorphic virus. Then, $\mathcal{L}_M(\mathscr{M}) \subseteq L\left(\mathcal{G}^S\right)$ for $\mathcal{G}$ a WG. To identify $L\left(\mathcal{G}^S\right)$ as a virus, an AVT must embed the Turing machine $\mathscr{M}$ s.t. $\mathcal{L}(\mathscr{M}) \in \mathcal{V}$.